\documentclass[12pt]{article}
\usepackage{epsfig}

\begin{document}

\begin{titlepage}

\hfill FTUV-11-2702

\hfill IFIC/11-14

\vspace{1.5cm}

\begin{center}
\ 
\\
{\bf\large Looking for magnetic monopoles at  LHC}
\\
\date{ }
\vskip 0.70cm

Luis N. Epele$^{a}$, Huner Fanchiotti$^{a}$, Carlos A. Garc\'{\i}a
Canal$^{a}$, Vasiliki A. Mitsou$^{b}$
\\ and Vicente Vento$^{b,c}$

\vskip 0.30cm

{(a) \it Laboratorio de F\'{\i}sica Te\'{o}rica, Departamento de
F\'{\i}sica, IFLP \\ Facultad de Ciencias Exactas, Universidad
Nacional de La Plata
\\C.C. 67, 1900 La Plata, Argentina.}\\({\small
E-mail: epele@fisica.unlp.edu.ar, huner@fisica.unlp.edu.ar, garcia@fisica.unlp.edu.ar})

\vskip 0.3cm
{(b) \it Instituto de F\'{\i}sica Corpuscular\\
 Universidad de Valencia and CSIC\\
Apartado de Correos 22085, E-46071 Valencia, Spain.}\\
({\small E-mail:
vasiliki.mitsou@ific.uv.es})\\
\vskip 0.3cm 
{(c) \it Departamento de F\'{\i}sica Te\'orica \\
Universidad de Valencia \\
E-46100 Burjassot (Valencia), Spain.} \\ ({\small E-mail:
vicente.vento@uv.es}) 
\end{center}

\vskip 1cm \centerline{\bf Abstract}

Magnetic monopoles have been a subject of interest since Dirac established the
relation between the existence of monopoles and charge quantization.
The intense experimental   search carried thus far has not met with success. 
We study the observability  of monopoles at the Large Hadron Collider  in
 the $\gamma \, \gamma$ channel and show that LHC is an ideal machine to
 discover monopoles with masses  below 1 TeV at present running energies and with less than 1~fb$^{-1}$
 of integrated luminosity.

 \vspace{1cm}

\noindent Pacs: 14.80.Hv, 95.30.Cq, 98.70.-f, 98.80.-k


\end{titlepage}

\section{Introduction}

The theoretical justification for the existence of classical magnetic poles, hereafter called
monopoles, is that they add symmetry to Maxwell's equations and explain charge 
quantization \cite{Dirac:1931kp,Jackson:1982ce}. Dirac showed that the mere existence 
of a monopole in the universe could
offer an explanation of the discrete nature of the electric charge. His analysis leads to the
 Dirac Quantization Condition (DQC),

\begin{equation} e \, g = \frac{N}{2} \;, \mbox{  N = 1,2,...}\;, 
\label{dqc}\end{equation}

\noindent where $e$ is the electron charge, $g$ the monopole
magnetic charge and we use natural units $\hbar = c =1$.
In Dirac's formulation, monopoles are assumed to exist as point-like particles and quantum mechanical
consistency conditions lead to Eq.(1), establishing the value of their magnetic charge.
Their mass, $m$, is a parameter of the theory.

Monopoles and their experimental detection
have been a subject of much study since many believe in Dirac's statement \cite{Dirac:1931kp},

\begin{center}
{\it ``...one would be surprised if Nature had made no use of it [the monopole]."}
\end{center}
\vskip 0.2cm
At present, despite intense experimental search, there is no evidence of their existence
\cite{Milton:2006cp,Yao:2006px}. LHC is opening a new frontier in the search
for monopoles and, as we show here, should  find them at present energies,  with less than 1~fb$^{-1}$ integrated 
luminosity, if their mass is below 1 TeV.

Although monopoles symmetrize Maxwell's equations in form there is a numerical
asymmetry arising from the DQC, namely that the basic magnetic charge is much larger
than the smallest electric charge. This led Dirac himself in his 1931 paper \cite{Dirac:1931kp} to state,

\begin{center}
\begin{minipage}{6in}
{\it ``... the attractive force between two one quantum poles of opposite sign is $( 137
2 )^2 \approx 46921 \frac{1}{4}$ times that between the electron and the proton. This very large force may perhaps
account for why the monopoles have never been separated."}
\end{minipage}
\end{center}
\vskip 0.2cm
This statement by Dirac is fundamental in our investigation. If instead of monopoles we deal with monopole-antimonopole 
pairs, we expect that due to this very strong interaction many of them annihilate into photons  
inside the detector.  We study  the production of a monopole-antimonopole pairs via photon fusion at LHC  and
 their subsequent annihilation giving rise to two extremely energetic photons. 

Another proposal,  influenced also by the strength of the interaction, assumes that the produced  
pair of monopole-antimonopole,  before decaying,  forms a monopolium bound state \cite{Epele:2007ic,Epele:2008un}. 
This bound state will also desintegrate producing two photons, a process which is being analyzed at present \cite{LaPlata2011}
in connection with LHC potentialities.

In the next section we describe the dynamics of monopoles and review the production of monopole-antimonopole pairs.
Section 3 discusses  the annihilation of the virtual monopole-antimonopole pair into photons. In section 4 we describe how to incorporate the elementary 
process into $p-p$ scattering. In section 5 we present our results in the context of the present running features of LHC and in the last section we draw conclusions of our studies.

\section{Monopole dynamics}

\begin{figure}[b]
\begin{center}
\includegraphics[scale= 0.6]{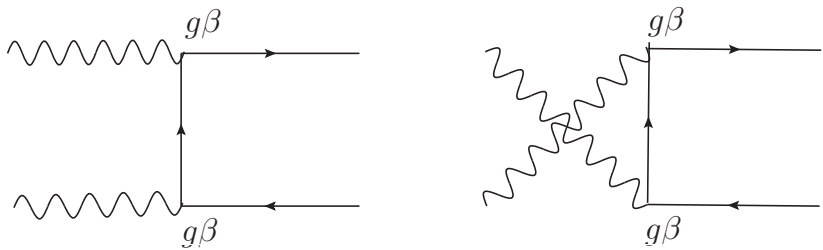}
\caption{ Elementary processes of monopole-antimonopole production via photon fusion.}
\label{mmproduction} 
\end{center}
\end{figure}

The theory of monopole interactions was initially formulated by Dirac \cite{Dirac:1948um} and later on developed in two different approaches
by Schwinger \cite{Schwinger:1966nj} and Zwanziger \cite{Zwanziger:1970hk}.  The formalism of Schwinger can be cast in 
functional form as a field theory for monopole-electron interaction which is dual to 
Quantum Electrodynamics (QED) \cite{Gamberg:1999hq}. The monopoles are considered fermions and behave as 
electrons in QED with a large coupling constant as a result of the DQC. In this formulation  the conventional photon 
field is Dirac string dependent. Due to the large coupling constant and the string 
dependence, perturbative treatments \`a la Feynman are in principle 
not well defined. However, non perturbative high energy treatments, like the eikonal approximation, 
have rendered well defined electron-monopole cross sections \cite{Gamberg:1999hq,Urrutia:1978kq}.

For the case of monopole production at energies higher than their mass the above procedure is not 
applicable, and being the treatment non perturbative, there is no universally accepted prediction 
from field theory.  However, the study of  the  classical interaction of a monopole passing by a 
fixed electron  leads to an interaction for the monopole which is associated with the electric field 
felt. If one compares this interaction with the that of  a positron passing by an electron, one realizes 
that the difference between QED and dual QED is simply to change the electric charge by the magnetic 
charge times the velocity, i.e.

\begin{equation}
e \rightarrow \beta  g .
\end{equation}
Thus a monopole behaves as a passing  particle with electric charge $\beta g$ \cite{Mulhearn:2004kw}.

This idea has been used to define an effective field theory  to lowest order \cite{Kalbfleisch:2000iz} which has been applied 
to Drell-Yan like monopole, or monopole-antimonopole production mechanisms \cite{Kalbfleisch:2000iz,Abulencia:2005kq} and to monopole-antimonopole 
production by photon fusion \cite{Kurochkin:2006jr,Dougall:2007tt}  . In Fig. \ref{mmproduction} we show the corresponding diagrams that have been calculated for the latter process.

We note that this approximation can be shown to be almost equivalent to the low
energy effective theory of Ginzburg and Schiller 
\cite{Ginzburg:1998vb,Ginzburg:1999ej}.
The theory of Ginzburg and Schiller  was derived from
the standard electroweak theory in the one loop approximation 
leading to an effective coupling  of the order of $ g_{eff} \sim
\frac{\varepsilon}{m}\, g$, where $m$ is the monopole mass and $\varepsilon$ a kinematical energy scale
of the process which is below the monopole production threshold, 
thus rendering the effective theory perturbative. In a
photon fusion diagram the dynamical scale is $\sqrt{E^2-4m^2}$, where $E$ is the center of mass energy, thus

\begin{equation}
\frac{\varepsilon}{m} \sim \frac{\sqrt{E^2-4m^2}}{2m}\sim \frac{E \beta
}{2m},
\end{equation}
where $\beta=\sqrt{1-\frac{4m^2}{E^2}}$ is the
monopole velocity.  Therefore if $E \sim 2m$, i.e. the kinetic energy is small, and  both
schemes coincide \cite{Epele:2008un} . 

The aim here is to study   possible  signals of magnectic monopoles at LHC. According to previous studies \cite{Dougall:2007tt}, the most promising mechanism is photon fusion. The elementary diagrams contributing to pair production are those in Fig. \ref{mmproduction}, where the explicit couplings have been shown. 


\begin{figure}[htb]
\begin{center}
\includegraphics[scale= 0.9]{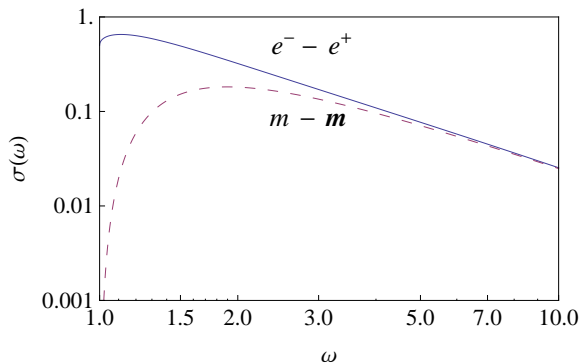}
\caption{ Elementary photon-fusion cross section for electron-positron (solid) and that of the monopole-antimonopole (dashed) as a function of $\omega$, to show the importance of the $\beta$ effect. } 
\label{mmxsection} 
\end{center}
\end{figure}


The photon-fusion elementary cross section  is obtained from the well-known QED electron-positron pair creation cross section \cite{Itzykson:1980rh}, simply changing the coupling constant  ($e \rightarrow g\beta$ ) and the eletron mass by the monopole mass $m_e \rightarrow m$, leading to

\begin{equation}
\sigma(\gamma \,\gamma \rightarrow m \overline{m})= \frac{\pi\;g^4
\,(1-\beta^2)\,\beta^4}{2 \,m^2}
\left(\frac{3-\beta^4}{2 \beta}
\log\left({\frac{1+\beta}{1-\beta}}\right) -(2-\beta^2)\right),
\label{ggxsecmm}
\end{equation}
where  $E$  the center of mass energy and $\beta$, a function of $E$, is the monopole velocity.  In Fig. \ref{mmxsection} we show the $\omega= E/2\, m$ dependence of Eq.(\ref{ggxsecmm}). The solid curve corresponds to the electron-positron case, the dashed one to the monopole case which contains the $\beta^4$ factor. One should notice the large effect associated with this factor in the vicinity of the threshold.

LHC detectors, apart from the MoEDAL experiment \cite{moedal},  have not been designed specifically to see monopoles and therefore even those which do not annihilate inside the detectors will be difficult to detect. However, all detectors are well designed to measure low cross sections of photons, since the two photon decay is one of the favorable channels to detect a low mass Higgs, thus we would like to benefit from this capability and proceed to investigate how monopole-antimonopole annihilation into photons is described in our scheme.

\section{Monopole-antimonopole annihilation into $\gamma \, \gamma$}


\begin{figure}[b]
\begin{center}
\includegraphics[scale= 0.6]{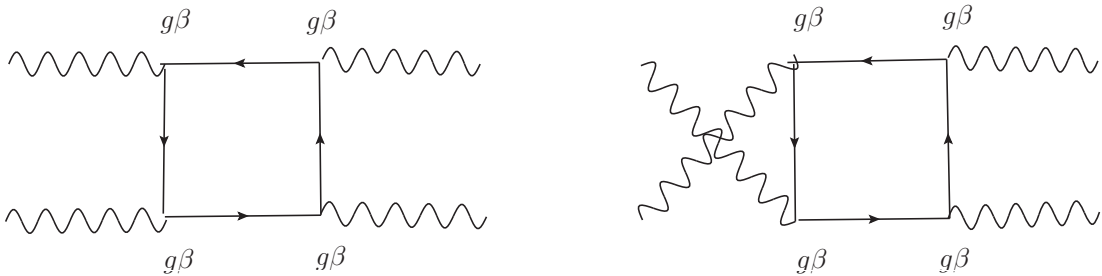}
\caption{ Elementary processes for  monopole-antimonopole production and annihilation into photons.}
\label{mmannihilation} 
\end{center}
\end{figure}


It is natural to think that the enormous strength and long range of the  monopole interaction leads to the annihilation of the pair  into photons very close to the production point. Thus one should  look for monopoles through their annihilation into highly energetic photons, a channel for which LHC detectors have been optimized.
 
In order to calculate the annihilation into photons we  assume that our effective theory, a technically convenient modification of Ginzburg and Schiller's, agrees with QED at one loop order,  and therefore we apply light-by-light scattering with the appropriate modifications as shown in Fig. \ref{mmannihilation}.
An interesting feature of the calculation is that the additional magnetic coupling will increase the cross section dramatically and therefore this measurement
should lead to a strong restriction on the monopole mass. However, as we have seen, in $m-\overline{m}$ production, the large magnetic coupling is always multiplied by the small electric one, leading to effective couplings  $e\, g$, and the same will happen in detection. Thus, the effective coupling of the process is $e \, g$, and therefore has strengths similar to the strong interaction, not more. The extreme sensitivity of LHC detectors to photons, as a consequence of the hope to detect the Higgs in the $\gamma \, \gamma $ channel makes them ideal for the purpose of detecting monopoles, even if the monopole mass is large, as will be shown.

\begin{figure}[htb]
\begin{center}
\includegraphics[scale= 0.9]{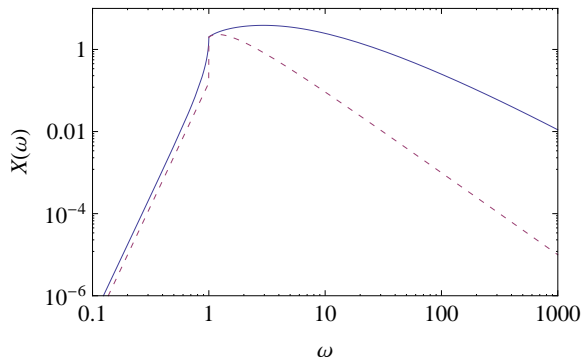}
\caption{The light-by-light scattering quantity $X(\omega)$, which determines the cross section up to factors related to the coupling constant. 
The solid curve corresponds to forward scattering, the dashed one to right-angle scattering, both as a function of the dimensionless quantity $\omega$.}
\label{lightlight} 
\end{center}
\end{figure}

Light-by-light scattering was studied  by Karplus and Neuman \cite{Karplus:1950zza,Karplus:1950zz}, who reproduced all previous low energy results by Euler \cite{Euler:1936zz} and high energy results by Achieser \cite{Achieser:1937zz}, and was later revisited and corrected by Csonka and Koelbig \cite{Csonka:1974ey}.

\begin{figure}[htb]
\begin{center}   

\includegraphics[scale= 0.75]{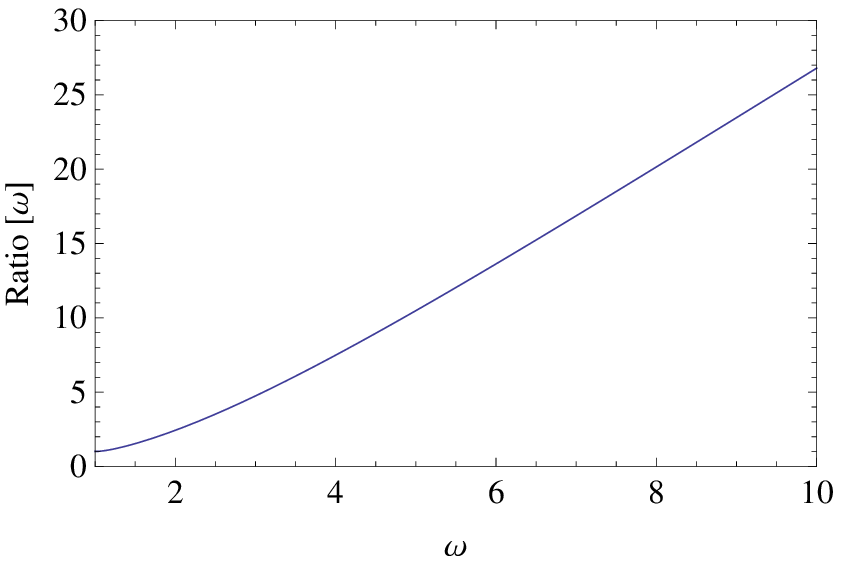}
\hskip 1cm
\includegraphics[scale= 0.75]{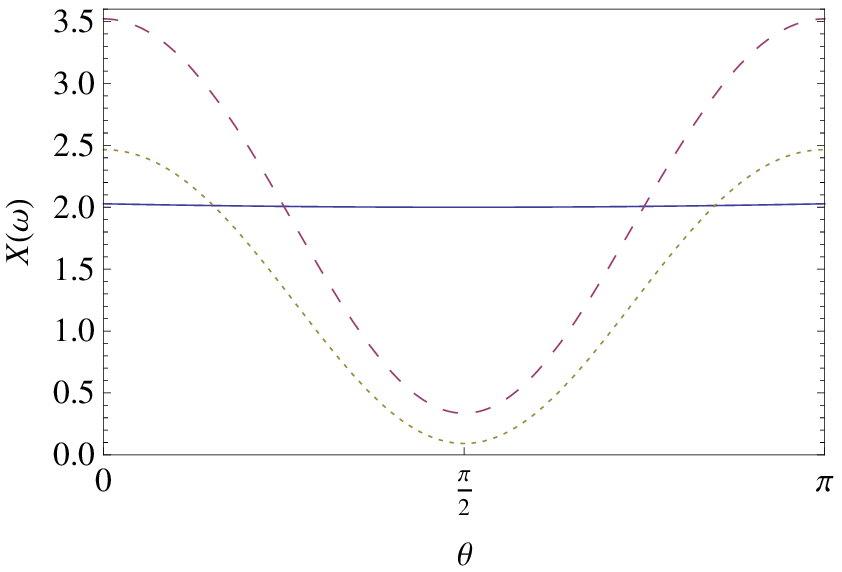}
\caption{Left: ratio of the  light-by-light scattering quantity $X(\omega)$ for forward to right-angle scattering as a function of $\omega$.  Right: angular dependence of the cross section for three values of $\omega$ (1 (solid) , 5 (dashed), 10 (dotted)) . }
\label{comparison} 
\end{center}
\end{figure}

In the case of monopoles, with the appropriate changes, the expression for the cross section becomes, 

\begin{equation}
 \sigma_{\gamma \gamma} (\theta, \omega) = \frac{(g \beta)^8}{8 \pi^2 m^2}  \; X(\theta , \omega).
\label{lightlightxsection}
\end{equation}
where $X(\omega)$ is $\frac{<|M|^2>}{\omega^2}$ in the notation of ref. \cite{Karplus:1950zz}.
In Fig. \ref{lightlight} we reproduce these results for forward scattering and right-angle scattering.  In Fig. \ref{comparison} (left)  we show the ratio of the forward scattering  to the right-angle  cross sections and therefore  show how  the cross section, which is basically isotropic close to  threshold, becomes anisotropic as the energy increases. In the same figure (right) we plot  $X(\omega)$ for several values of $\omega$  as a function of angle. We note that the forward cross section is larger and the right-angle one  smaller, than that for any other value of the scattering angle. As the energy increases the drop in the $X$ function from $\theta= 0$ to $\pi/2$ increases.

 It is clear from these figures that close to the threshold the cross section is quite isotropic and  away from threshold the forward cross section, which is very difficult to measure, is much larger than the right-angle one.  Since the detectors cannot detect all of the photons coming out, we  take the right-angle cross section as a conservative indication of the magnitudes to be expected.

\begin{figure}[htb]
\begin{center}   
\includegraphics[scale=0.9]{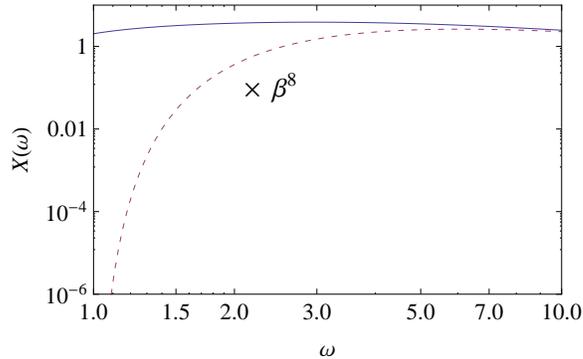}
\caption{The effect of the $\beta$ factor  as a function of $\omega$. The solid curve corresponds to electron scattering, while the dashed curve corresponds to 
the monopole case. }
\label{beta} 
\end{center}
\end{figure}

In Fig. \ref{beta} we show the effect of the $\beta$ factor which diminishes greatly the cross section close to the monopole-antimonopole threshold.

\section{Analysis of p-p scattering}

\begin{figure}[t]
\begin{center}   
\includegraphics[scale=0.6]{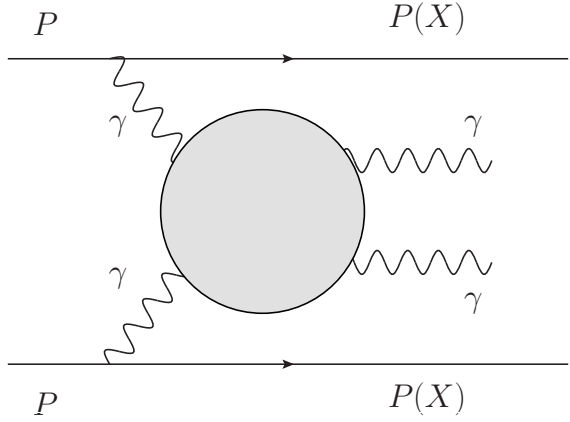}
\caption{Processes contributing to the $\gamma \gamma$ cross section. The blob contains the three cases described in the text. }
\label{pp} 
\end{center}
\end{figure}

LHC is a proton-proton collider, therefore, in order to describe the production and desintegration of the monopole-antimonopole pair, 
we have to study the following proceses above the monopole threshold ($ \beta > 0 \rightarrow E \geq 2 m$),

\begin{eqnarray}
p + p &  \rightarrow &  p(X) + p(X) + \gamma + \gamma,
\end{eqnarray}
shown globally in Fig.\ref{pp}, where $p$ represents the proton, $X$ an
unknown final state and we will assume that the blob is due exclusively to monopoles. This diagram summarizes
the three possible processes:

\begin{itemize}

\item [i)] inelastic $p+ p \rightarrow X+X + \gamma +\gamma
\rightarrow X + X + m + \overline{m} \rightarrow  X + X + m + \overline{m} +\gamma + \gamma$

\item [ii)] semi-elastic $ p + p \rightarrow p + X + \gamma + \gamma
\rightarrow p + X +  m + \overline{m}  \rightarrow  p + X + m + \overline{m} +\gamma + \gamma$

\item [iii)] elastic $p + p \rightarrow p + p + \gamma + \gamma
\rightarrow p + p +  m + \overline{m} \rightarrow \rightarrow  X + X + m + \overline{m} + \gamma + \gamma$.
\end{itemize}

In the inelastic scattering, both intermediate photons are radiated
from partons (quarks or  antiquarks) in the colliding protons.

In the semi-elastic scattering one intermediate photon is radiated
by a quark (or antiquark), as in the inelastic process, while the
second photon is radiated from the other proton, coupling to the
total proton charge and leaving a final state proton intact.

In the elastic scattering both intermediate photons are radiated
from the interacting protons leaving both protons intact in the
final state.

In the blob we incorporate the elementary subprocess shown in Fig.
\ref{mmannihilation} and  described by Eq. (\ref{lightlightxsection}).

We calculate $\gamma \gamma$ fusion for monopolium-antimonopolium production
following the formalism of Drees et al. \cite{Drees:1994zx}.

In the inelastic scattering, $p + p\rightarrow X+ X + (\gamma
\gamma) \rightarrow X +X + m+ \overline{m} + \gamma + \gamma$, to approximate the quark distribution
within the proton we use the Cteq6--1L parton distribution functions
\cite{CTEQ} and choose $Q^2=\hat{s}/4$ throughout.

We employ an equivalent--photon approximation for the photon
spectrum of the intermediate quarks \cite{Williams:1934ad,von
Weizsacker:1934sx}.

In semi--elastic scattering, $p + p\rightarrow p+ X+ (\gamma \gamma)
\rightarrow p+ X + m+ \overline{m} + \gamma + \gamma $, the photon spectrum associated with the
interacting proton must be altered from the equivalent--photon
approximation for quarks to account for the proton structure.  To
accommodate the proton structure we use the modified
equivalent--photon approximation of \cite{Drees:1994zx}.

The total cross section is obtaned 
as a sum of the three processes. The explicit expressions for the
different contributions can be found  in \cite{Dougall:2007tt}.

In order to obtain the differential photon-photon cross section from the above
formalism we develop a procedure  which we exemplify with the elastic 
scattering case. In that case the $pp$ cross section is given by \cite{Drees:1994zx},

\begin{equation}
\sigma_{pp} (s)  = \int_{4 m^2/s}^1 d z_1  \int_{4 m^2/s z_1}^1 d z_2 f(z_1) f(z_2) \sigma_{\gamma \gamma} (z_1 z_2  s),
\end{equation}
where $\sqrt{s}$ is the center of mass energy of the $pp$ system.

We perform the following change of variables 
$$ v= z_1 z_2  \;  , \; w= z_2 \; ,$$
which leads to 
\begin{equation}
\sigma_{pp} (s)  = \int_{4 m^2/s}^1 d v  \int_{v}^1 \frac{d w}{w}  f(\frac{v}{w}) f(w) \sigma_{\gamma \gamma} (v s). 
\end{equation}
Note that to fix the center of mass energy of the photons is equivalent to fix $v$. For fixed $v$ we have,

\begin{equation}
\frac{d \sigma_{pp}}{dv} (s)  = \int_{v}^1  \frac{d w}{w}  f(\frac{v}{ w}) f(w) \sigma_{\gamma \gamma} (v s),
\end{equation}
which can be rewritten in terms  of $E_\gamma$, the center of mass energy  of the photons, and the elementary photon-photon cross section as,

\begin{equation}
 \frac{d \sigma_{pp}}{dE} (E_\gamma) = \frac{2 E_\gamma}{s}  \; \sigma_{\gamma \gamma} (s_{\gamma \gamma})\;\int_{s_{\gamma  \gamma}/s}^1  \frac{d w}{w}  f(\frac{s_{\gamma  \gamma}}{ w}) f(w).  
\end{equation}
This procedure can be generalized easily to the semielastic and inelastic cases, where the appropriate change of variables are

$$ v= z_1 z_2 x_1  \; , \;  w= z_2 x_1 \; , \;  u =  x_1$$
and
$$ v= z_1 z_2 x_1  x_2 \; , \;  w= z_2 x_1 x_2 \; , \;  u =  x_1 x_2 \; , \; t = x_2,$$
respectively.

\begin{figure}[t]
\begin{center}   
\includegraphics[scale=0.9]{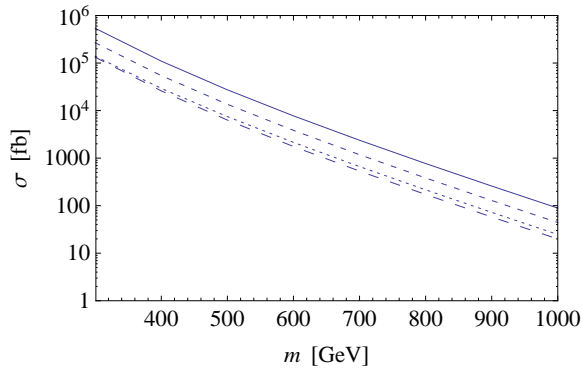}
\caption{The total monopole-antimonopole production cross section (solid line) from $\gamma \, \gamma$ fusion in fb as a function of monopole mass.  The different components of the cross section are shown: elastic (dotted line), semielastic (small dashed line) and inelastic (dashed line). }
\label{mmproductionxsec} 
\end{center}
\end{figure}

\section{Results}

In Fig. \ref{mmproductionxsec} we show the total cross section for monopole-antimonopole  production and the contribution 
from each of the individual processes described above as a function of monopole mass. In order to set the mass axes we have taken into account the previous 
lower mass limit for the monopole of ref. \cite{Abulencia:2005kq} which was set at $360$ GeV. The cross section is of $\mathcal{O}({\rm fb})$ thus we limit the high mass values to those potentially observable at present, or in the near future, by the LHC with an integrated luminosity of 1~fb$^{-1}$.
 Our result differs from that of ref. \cite{Dougall:2007tt} around threshold.

We see in the previous figure that monopoles of mass around 500 GeV should be produced abundantly, while those of mass around 1000 GeV have a cross section which makes them difficult to detect in the near future, at least directly.

\begin{figure}[htb]
\begin{center}   
\includegraphics[scale=0.9]{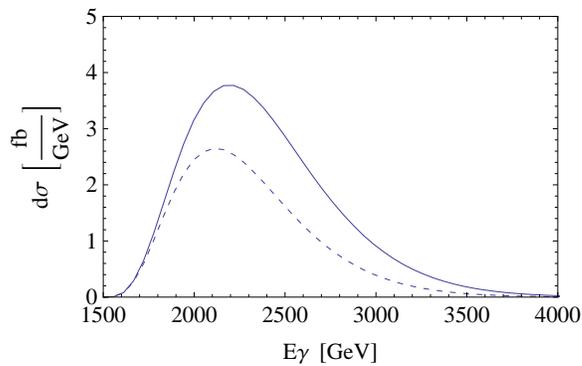}
\caption{Forward (solid) and the right-angle cross sections for a monopole of mass $750$ GeV.}
\label{0pi2} 
\end{center}
\end{figure}

Since, as already mentioned, the LHC detectors have not been tuned to detect monopoles but are excellent detectors for $\gamma$'s let us discuss next our annihilation cross section.  In Fig. \ref{0pi2} we show the differential cross section for forward (solid) and right-angle (dotted) scattering  given in  fb/GeV as a function of the invariant mass of the $\gamma\, \gamma$ system. We have assumed a monopole of mass $750$ GeV, chosen because the cross sections turned out to be close  to the expected magnitude of the  Higgs to $\gamma \; \gamma$ cross section above background. The cross sections are wide, almost gaussian, structures rising softly just above threshold ($1500$ GeV). LHC detectors are blind for forward scattering and have black spots due to construction features in the non-forward regions which do not allow for a full detection of photons. Therefore in order to obtain an educated estimation of  the observable cross section we  take  the right-angle  cross section and multiply it by $4 \pi$. This differential cross section is the smallest possible. However, away from threshold, it corresponds quite well to a realistic estimate, since, as we have seen, the elementary differential cross section drops fast with angle and moreover one should consider an efficiency factor for the various detectors.

\begin{figure}[htb]
\begin{center}   
\includegraphics[scale=0.9]{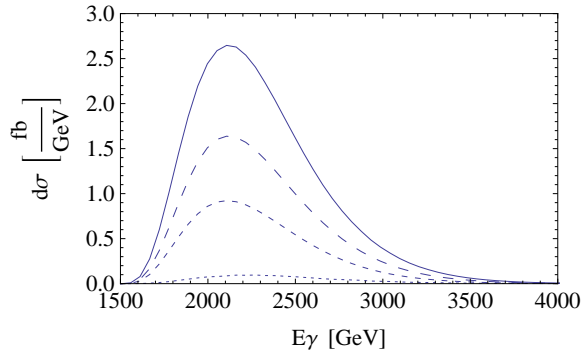}
\caption{The right-angle scattering cross sections for a monopole mass of $750$ GeV.  The smallest is the inelastic cross section (dotted), next comes the elastic (dashed) and the biggest is the semielastic (long-dashed). The total cross section, the sum of the three, is represented by the solid curve.}
\label{pi2} 
\end{center}
\end{figure}

 In Fig. \ref{pi2} we plot the right-angle differental cross section  for the same monopole mass and show the different contributions to the differential cross section. The structure of the cross section is a wide structure, rising softly after threshold, $1500$ GeV, and extending for almost $2000$ GeV. The structure is centered about $2300$ GeV. Clearly the soft rise of the differential cross section is a signature of the two particle threshold, recall the $\beta$ factor. The width of the stucture is associated to the mathematical form of the box diagram, as can be seen for both electron-positron annihilation and monopole-antimonopole, from the structue of the elementary cross section (Fig. \ref{beta}).

\begin{figure}[htb]
\begin{center}   
\includegraphics[scale=0.75]{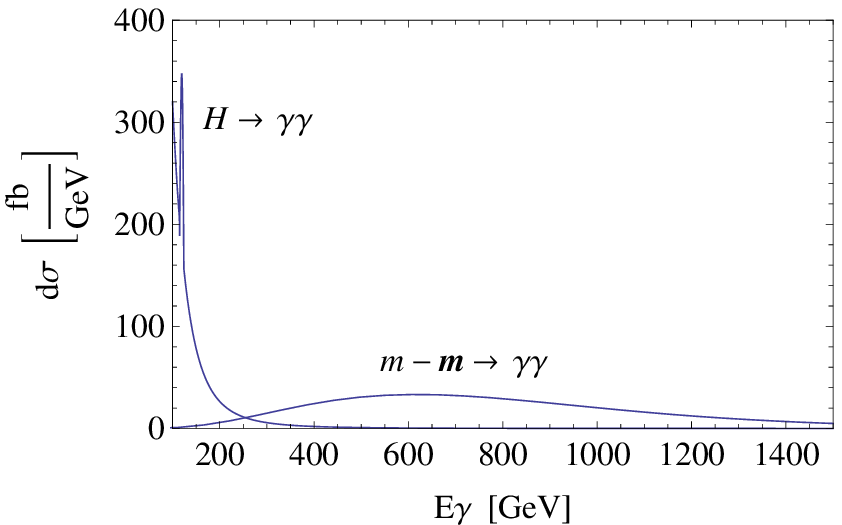}
\hskip 1.0cm
\includegraphics[scale=0.80]{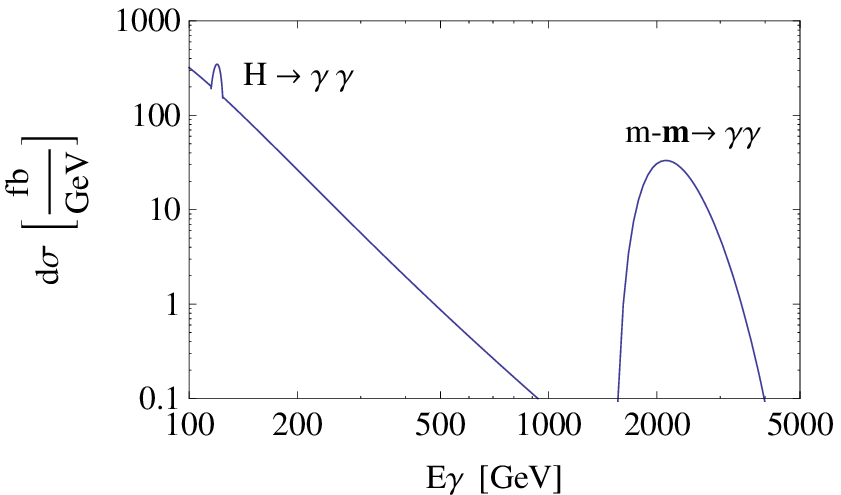}
\caption{Comparison of the $\gamma \gamma$ monopole-antimonopole annihilation cross section for a monopole of mass $750$ GeV  with the Higgs $\gamma \gamma$ decay. The Higgs cross section above the background has been multiplied by $50$. The monopole-antimonopole threshold has been set at the origin ($100$ GeV) (left). The right figure represents the same cross section drawn in a Log Log plot keeping the thresholds.}
\label{higgsmonopole} 
\end{center}
\end{figure}

In Fig. \ref{higgsmonopole} we compare our total $\gamma \gamma$ cross section with Higgs process obtained from ref. \cite{Aad:2010qr}. We have extrapolated their background to our energies using an inverse polynomial fit to their data and their exponential fit. Both procedures give a negligible background for the signals obtained  with monopoles masses up to 1 TeV and even higher. In the left figure we  traslate the monopole-antimonopole threshold to the origin in order to compare the two signals. In the right figure we show the actual energy scales in a LogLog plot. The Higgs signal above background has been multiplied by 50 to make it visible. The figures correspond to a monopole of mass $750$ GeV. It is clear from the curves that monopole-antimonopole annihilation should appear as a soft rise of the cross section  above the background over a large energy interval.

\begin{figure}[htb]
\begin{center}   
\includegraphics[scale=0.9]{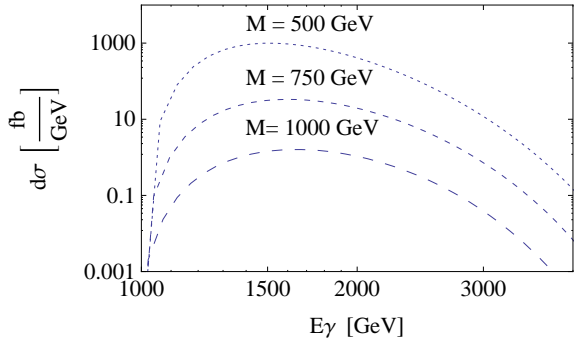}
\caption{The cross section for monopole-antimonopole annihilation into $\gamma \gamma$ for three masses: $500$ GeV (solid), $750$ GeV dashed $1000$ GeV (dotted).}
\label{masses} 
\end{center}
\end{figure}

Finally we study the mass dependence of the differential cross section. Fig. \ref{masses} is a LogLog plot of the cross section for three masses, $500, 750$ and $1000$ GeV. In order to have a better comparison we have translated the thresholds to the origin. It is clear that the magnitude of the cross section and its extent falls rapidly with mass. For a low mass monopole, the  magnitude of the differential cross section is of the order of  $\sim 1$ pb/GeV, for an intermediate mass monopole the cross section is of the order of   $\sim 10$ fb/GeV  and  for a heavy  monopole is of the order of $\sim 1$ fb/GeV.  If the monopole has a low mass, monopoles should be seen within the initial period of LHC running.

\section{Conclusions}

The existence of Dirac monopoles would modify our understanding of QED. The DQC, implying a huge magnetic coupling constant, complicates matters from the theoretical point of view. Non perturbative methods are required. We have avoided the problem by using an effective theory valid at one loop order. The physical coupling turns out to be $e\, g$, equivalent in magnitude to the pionic couplings, and the expected high monopole masses provide convenient cut-offs which make the theory sensible.  In the context of this theoretical scheme we are able to calculate monopole(antimonopole) production and monopole-antimonopole annihilation. Centering our description to LHC, a proton machine,  we have described scenarios for the production of monopole-antimonopole pairs via photon fusion. We have described their annihilation into photons via the conventional  box diagram, analogous to that of light-by-light scattering in conventional QED.  The only difference in our approach is the introduction of a threshold factor in the form of a velocity. We have analyzed in detail the elementary process, light-by-light scattering at monopole energies, and have subsequently incorporated the description of the photon-photon elementary scattering into the initial proton-proton collision. One main assumption is that no other particle contributes to the box diagram in the region of interest. We have chosen in the present calculation a mass range which is above conventional particle masses and below the supersymmetric particle ranges. One could think of interference with  elementary particle decays, however the latter would have a resonant structure which our box diagram does not have. Thus any interference would be avoided by their different geometric structure.

With all these preventions we have shown that monopoles up to 1 TeV in mass should be detected at LHC in the $\gamma \; \gamma$ channel at present running energies and with luminosities in reach within the next few months. The signature is a soft rise of the cross section which should stay for a long energy range, because no resonant peak structure is to be expected.

\section*{Acknowledgement}
We thank the authors of JaxoDraw  for making drawing diagrams an
easy task \cite{Binosi:2003yf}.  LNE, HF
and CAGC were partially supported by CONICET and ANPCyT Argentina. VAM acknowledges support by the Spanish Ministry of Science and Innovation (MICINN) under the project FPA2009-13234-C04-01, by the Spanish Agency of International Cooperation for Development under the PCI projects A/023372/09 and A/030322/10 and by the CERN Corresponding Associate Programme. VV has been supported  by HadronPhysics2,  by  MICINN (Spain) grants FPA2008-5004-E,  FPA2010-21750-C02-01,   AIC10-D-000598 and by GVPrometeo2009/129.


\begin{thebibliography}{60}


\bibitem{Dirac:1931kp}
  P.~A.~M.~Dirac,
  Proc.\ Roy.\ Soc.\ Lond.\  A {\bf 133} (1931) 60.
  
  
\bibitem{Jackson:1982ce} J. D. Jackson, Classical Electrodynamics,
de Gruyter, N.Y. (1982).


\bibitem{Milton:2006cp}
  K.~A.~Milton,
  Rept.\ Prog.\ Phys.\  {\bf 69} (2006) 1637
  [arXiv:hep-ex/0602040].


\bibitem{Yao:2006px}
  W.~M.~Yao {\it et al.}  [Particle Data Group],
  J.\ Phys.\ G {\bf 33} (2006) 1.




\bibitem{Epele:2007ic}
  L.~N.~Epele, H.~Fanchiotti, C.~A.~Garcia Canal and V.~Vento,
  Eur.\ Phys.\ J.\  C {\bf 56} (2008) 87
  [arXiv:hep-ph/0701133].
  
\bibitem{Epele:2008un}
  L.~N.~Epele, H.~Fanchiotti, C.~A.~G.~Canal and V.~Vento,
  Eur.\ Phys.\ J.\  C {\bf 62} (2009) 587
  [arXiv:0809.0272 [hep-ph]].
 
  \bibitem{LaPlata2011}
  L.~N.~Epele, H.~Fanchiotti, C.~A.~G.~Canal, V.~Mitsou and V.~Vento,
 (work in preparation).



\bibitem{Dirac:1948um}
  P.~A.~M.~Dirac,
  Phys.\ Rev.\  {\bf 74} (1948) 817.

\bibitem{Schwinger:1966nj}
  J.~S.~Schwinger,
  Phys.\ Rev.\  {\bf 144} (1966) 1087.

\bibitem{Zwanziger:1970hk}
  D.~Zwanziger,
  Phys.\ Rev.\  D {\bf 3} (1971) 880.

\bibitem{Gamberg:1999hq}
  L.~P.~Gamberg and K.~A.~Milton,
  Phys.\ Rev.\  D {\bf 61} (2000) 075013
  [arXiv:hep-ph/9910526].

\bibitem{Urrutia:1978kq}
  L.~F.~Urrutia,
  Phys.\ Rev.\  D {\bf 18} (1978) 3031.


\bibitem{Mulhearn:2004kw}
  M.~J.~Mulhearn,
  ``A Direct Search for Dirac Magnetic Monopoles,''
 Ph.D. Thesis, Massachusetts Institue of Technology 2004, FERMILAB-THESIS-2004-51.
 
 

  
  
\bibitem{Kalbfleisch:2000iz}
  G.~R.~Kalbfleisch, K.~A.~Milton, M.~G.~Strauss, L.~P.~Gamberg, E.~H.~Smith and W.~Luo,
  Phys.\ Rev.\ Lett.\  {\bf 85} (2000) 5292
  [arXiv:hep-ex/0005005].
  
 \bibitem{Abulencia:2005kq}
  A.~Abulencia {\it et al.}  [CDF Collaboration],
  Phys.\ Rev.\ Lett.\  {\bf 96} (2006) 011802
  [arXiv:hep-ex/0508051].

\bibitem{Kurochkin:2006jr}
  Yu.~Kurochkin, I.~Satsunkevich, D.~Shoukavy, N.~Rusakovich and Yu.~Kulchitsky,
  Mod.\ Phys.\ Lett.\  A {\bf 21} (2006) 2873.

\bibitem{Dougall:2007tt}
 T.~Dougall and S.~D.~Wick,
  Eur.\ Phys.\ J.\  A {\bf 39} (2009) 213
  [arXiv:0706.1042 [hep-ph]].

\bibitem{Ginzburg:1998vb}
  I.~F.~Ginzburg and A.~Schiller,
  Phys.\ Rev.\  D {\bf 57} (1998) 6599
  [arXiv:hep-ph/9802310].

\bibitem{Ginzburg:1999ej}
  I.~F.~Ginzburg and A.~Schiller,
  Phys.\ Rev.\  D {\bf 60} (1999) 075016
  [arXiv:hep-ph/9903314].

 
\bibitem{Itzykson:1980rh}
  C.~Itzykson and J.~B.~Zuber,
{\it  New York, USA: McGraw-Hill (1980) 705 P.(International Series In Pure and Applied Physics)}

\bibitem{moedal}
J.~L.~Pinfold,
  AIP Conf.\ Proc.\  {\bf 1304} (2010) 234;
J.~L.~Pinfold  [MOEDAL Collaboration],
  CERN Cour.\  {\bf 50N4} (2010) 19.


\bibitem{Karplus:1950zza}
  R.~Karplus and M.~Neuman,
  Phys.\ Rev.\  {\bf 80} (1950) 380.
  
\bibitem{Karplus:1950zz}
  R.~Karplus and M.~Neuman,
  Phys.\ Rev.\  {\bf 83} (1951) 776.

\bibitem{Euler:1936zz}
  H. Euler
  Ann.\ Phys.\  {\bf 26} (1936) 398.
 
 
\bibitem{Achieser:1937zz}
 A. Achieser,
  %
  Physik\ Z.\ Sowjetunion\  {\bf 11} (1937) 263.
 
\bibitem{Csonka:1974ey}
  P.~L.~Csonka and K.~S.~Koelbig,
  Phys.\ Rev.\  D {\bf 10} (1974) 251.


\bibitem{Drees:1994zx}
  M.~Drees, R.~M.~Godbole, M.~Nowakowski and S.~D.~Rindani,
  Phys.\ Rev.\  D {\bf 50} (1994) 2335
  [arXiv:hep-ph/9403368].

\bibitem{CTEQ} The Cteq6 parton distribution functions can be found
at

[http://www.phys.psu.edu//cteq/].


\bibitem{Williams:1934ad}
  E.~J.~Williams,
  Phys.\ Rev.\  {\bf 45} (1934) 729.

\bibitem{von Weizsacker:1934sx}
  C.~F.~von Weizsacker,
  Z.\ Phys.\  {\bf 88} (1934) 612.

\bibitem{Aad:2010qr}
  G.~Aad {\it et al.}  [Atlas Collaboration],
  Phys.\ Rev.\ Lett.\ {\bf 106} (2011) 121803  
  arXiv:1012.4272 [hep-ex].


\bibitem{Binosi:2003yf}
  D.~Binosi and L.~Theussl,
  Comput.\ Phys.\ Commun.\  {\bf 161} (2004) 76
  [arXiv:hep-ph/0309015].


\end{thebibliography}
\end{document}